\documentclass[aps,amsmath,amssymb,twocolumn,showpacs]{revtex4-1}

\usepackage{graphicx}
\usepackage{amsmath}
\usepackage{bm}
\usepackage{amssymb}
\usepackage{color}
\usepackage[utf8]{inputenc}

\newcommand{\sign}{\text{sign}}

\newcommand{\hb}[1]{\hat{\bar{#1}}}

\begin{document}

\title{Local density of states in clean two-dimensional superconductor--normal metal--superconductor heterostructures}

\author{D. Nikoli\' c}
\affiliation{Fachbereich Physik, University of Konstanz, D-78467 Konstanz, Germany}

\author{J.C. Cuevas}
\affiliation{Departamento de F\'{\i}sica Te\'orica de la Materia Condensada
and Condensed Matter Physics Center (IFIMAC), Universidad Aut\' onoma de Madrid, 
E-28049 Madrid, Spain}

\author{W. Belzig}
\affiliation{Fachbereich Physik, University of Konstanz, D-78467 Konstanz, Germany}

\begin{abstract}

Motivated by recent advances in the fabrication of Josephson junctions in which the weak link
is made of a low-dimensional non-superconducting material, we present here a systematic
theoretical study of the local density of states (LDOS) in a clean 2D normal metal (N) coupled to 
two s-wave superconductors (S). To be precise, we employ the quasiclassical theory of superconductivity
in the clean limit, based on Eilenberger's equations, to investigate  the 
phase-dependent LDOS as function of factors such as the length or the width of the junction, a finite 
reflectivity, and a weak magnetic field. We show how the the spectrum of Andeeev bound states that appear inside the gap shape the phase-dependent LDOS in short and long junctions. We discuss the circumstances when a gap appears in the LDOS and when the continuum displays a significant phase-dependence. The presence of a magnetic flux leads to a complex interference behavior, which is also reflected in the supercurrent-phase relation. Our results agree qualitatively with recent experiments on graphene SNS junctions. Finally, we show how the LDOS is  connected 
to the supercurrent that can flow in these superconducting heterostructures and present an analytical 
relation between these two basic quantities.

\end{abstract}

\date{\today}

\maketitle

\section{Introduction} \label{introduction}

If a normal metal (N) is in good electrical contact with a superconductor (S), it can acquire
genuine superconducting properties. This phenomenon, which is known as \emph{proximity
effect}, was first investigated in the 1960's \cite{deGennes1964,Deutscher1969} and there has been
a renewed interest in the last decades due to the possibility to study this effect at much smaller
length and energy scales \cite{Pannetier2000} and in novel low-dimensional materials. The 
proximity effect manifests itself in a modification of the LDOS of the normal metal and it 
is mediated by the so-called Andreev reflection \cite{Andreev1964}. In this process, an
electron of energy $E < \Delta$, where $\Delta$ is the superconducting gap in S, inside the 
normal metal impinges in the SN interface and is reflected as a hole transferring a Cooper pair 
to the S electrode. When the normal metal is sandwiched between two superconducting leads, 
multiple Andreev reflections can occur at the SN interfaces leading to the formation of 
Andreev bound states (ABSs) inside the gap region \cite{Kulik1970}. These ABSs are, in turn,
largely responsible for the supercurrent that can flow through the SNS junction when there is
a finite superconducting phase difference between the superconducting leads \cite{Kulik1970}.

In the last 50 years the Josephson effect in SNS weak links has been thoroughly investigated in 
numerous experiments in which the normal link ranged from standard normal metals
\cite{Tinkham1996,Barone1982,Clarke1969,Shepherd1972,Dubos2001} to low-dimensional materials 
such as carbon nanotubes \cite{Kasumov1999}, semiconductor nanowires \cite{Doh2005} or graphene 
\cite{Heersche2007}, just to mention a few. However, experimental studies exploring the 
LDOS in a normal metal in proximity to a superconductor are more scarce and they have only
been reported in recent years. The proximity-induced modification of the LDOS has been
spatially resolved with the help of local tunneling probes \cite{Gueron1996,Belzig1996,Meschke2011} 
and by means of Scanning Tunneling Spectroscopy (STS) technique applied to mesoscopic 
systems \cite{Chapelier2001,Moussy2001,Escoffier2004,leSueur2008,Wolz2011}. This method has been further refined to  probe the proximity effect in 2D metals with high spatial and energy resolution \cite{Kim2012,Serrier-Garcia2013,Cherkez2014,Roditchev2015}. These experiments
have been successfully explained with the help of the quasiclassical theory of superconductivity and
the so-called Usadel equations \cite{Usadel1970}, which describes the proximity effect in the 
dirty limit, i.e., when the elastic mean free path is much smaller than the superconducting
coherence length in the normal region. In another context, the local density of states has been probed in ferromagnetic proximity systems in order to probe the pairing symmetries. For instance, a zero-bias peak in the density of states relates to a mixed-spin triplet pairing \cite{Huertas2002,Linder2009,Linder2010,DiBernardo2015} or a triplet gap related to equal-spin Cooper pairing \cite{Diesch2018}.

In the regime, known as the clean limit, the mean free path is larger than both the junction and the coherence length. The LDOS is expected to display discrete ABSs inside the gap \cite{Kulik1970,Furusaki1991,Beenakker1991}.
To our knowledge, these discrete ABSs have only been resolved with tunneling probes in SNS 
heterostructures based on normal quantum dots, i.e., 0D systems, formed in carbon nanotubes 
\cite{Pillet2010}, graphene \cite{Dirks2011} or semiconductor nanowires \cite{Chang2013}. 
A natural candidate to observe ABSs in a 2D clean metal is graphene. In fact, the proximity 
superconductivity in graphene systems has been intensively investigated since its early days \cite{Titov2007} and has recently been reviewed in Ref.~\cite{Lee2018}. Remarkably, a two-dimensional interference pattern has been predicted in warped Fermi-surface proximitized in two-dimensions \cite{Ostroukh2016} or in the presence of boundaries \cite{Meier2016}.

In a recent work Bretheau and coworkers \cite{Bretheau2017} used a van der Waals heterostructure 
to perform tunneling spectroscopy measurements of the proximity effect in 
superconductor-graphene-superconductor junctions. By incorporating these heterojunctions in
a superconducting loop, they were able to measure the phase-dependent DOS in the graphene
region. Due to the large width of the junction they reported a continuum of ABSs, which clearly indicates that the junctions were not strictly in the one-dimensional limit, these experiments demonstrated the 
feasibility to fabricate and investigate clean 2D SNS junctions. Interestingly, the
authors of that work also postulated a heuristic relation between the supercurrent and the LDOS,
which allowed them to establish a infer the current-phase relation from their LDOS measurements. 

The LDOS and the corresponding ABS spectrum in clean 3D SNS junctions have been discussed earlier in the literature. The impact on the ABS spectrum of non perfect interfaces \cite{Nagato1993,Hara1993} and the possible pairing in the normal metal \cite{Nagato1995,Tanaka1991} by employing a self-consistent treatment of the pair potential in quasi-onedimensional SNS junctions have been studied. The authors of Ref.~\cite{Wendin1996} considered the proximity effect in a S-2DEG-S junction in the short junction limit.  However, a systematic theoretical analysis of the  LDOS  in junctions of arbitrary sizes and non perfect transparency with and without magnetic field has not be done so far.

In this Article, motivated by the experiments of Bretheau \emph{et al.}\ \cite{Bretheau2017}, we present a systematic study of the LDOS in clean 2D SNS junctions. We will make use of the
quasiclassical theory of superconductivity in the clean limit, which is based on the 
so-called Eilenberger equations \cite{Eilenberger1968}, to study the impact on the
phase-dependent LDOS of parameters such as the junction length and 
width, the transmission of the system, and the presence of a weak magnetic field. The use
of the quasiclassical Green's function formalism allows us to determine not only the
ABSs, but also the phase dependence of continuum of states outside the gap region. Moreover, 
we revisit the relation between LDOS and supercurrent proposed in Ref.~\cite{Bretheau2017} and 
show that that the correct formula should relate the supercurrent density to the global density of states, which leads to significant changes in the limit of relatively short junctions.  

The rest of the paper is organized as follows. In Sec.~\ref{system_method} we introduce the system
under study and describe in detail the quasiclassical Green's function formalism that we employ to
compute the LDOS in clean 2D SNS junctions. In particular, we discuss in different subsections
how to compute the LDOS in a fully transparent junction, how to account for the presence of a
potential barrier in the systems and how to describe the role of a finite width of the normal
region and the presence of a weak magnetic field. Section \ref{results} is devoted to the 
description of the main results of this work. In this section we illustrate the impact of different
factors, such as the length, the barrier transmission or the presence of a weak magnetic field, in
the LDOS in the normal region of a clean SNS junction. In Sec.~\ref{discussion} we present a 
discussion of the magnetic-field modulation of the LDOS in close connection to the work of 
Ref.~\cite{Bretheau2017} and present an analytical relation between LDOS
and the supercurrent in fully transparent junctions. Finally, Sec.~\ref{conclusion} contains 
a summary of our main results and conclusions.

\section{System and Method} \label{system_method}

\begin{figure}[t]
\includegraphics[width=\columnwidth,clip]{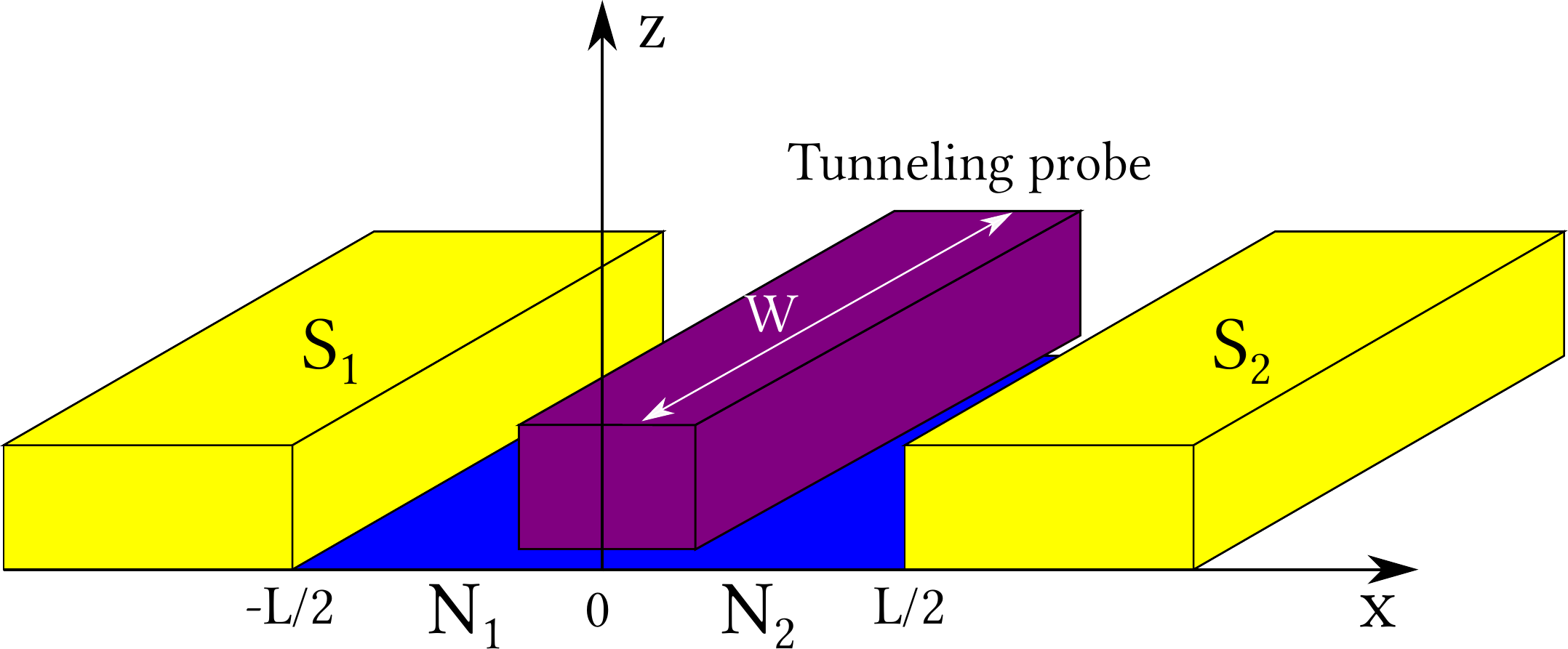}
\caption{(color online) Schematic representation of the system under study where a clean 2D normal metal
of length $L$ and width $W$ is coupled to two s-wave superconducting electrodes. The additional electrode
that appears on top of the normal region represents an eventual tunneling probe that could be used to 
measure the local DOS in the normal metal.}
\label{fig-system}
\end{figure}

Our goal is to calculate the local DOS in a clean (no impurities) 2D normal metal sandwiched by two identical 
s-wave superconductors, see Fig.~\ref{fig-system}. We assume the normal region to have a length $L$ and a 
width $W$. Eventually, we shall consider the role of interface scattering by considering the presence of 
a potential barrier in the middle of the normal metal characterized by a transmission coefficient $D$ that 
takes values from 0 to 1. In what follows, all the energy scales will be expressed in units of the superconducting
energy gap of the electrodes, $\Delta$, and the lengths will be compared to the superconducting
coherence length (inside the normal metal), which in the clean limit is given by $\xi = \hbar v_{\rm F}/\Delta$,
where $v_{\rm F}$ is the magnitude of the Fermi velocity. Moreover, in the following discussion we shall 
set $\hbar = 1$ and $k_{\rm B} = 1$ in most calculations, but reinsert them in selected final results.

In order to describe the electronic structure in this SNS heterostructure we make use of the quasiclassical
theory of superconductivity, which in the clean case is based on the Eilenberger equation of motion 
\cite{Eilenberger1968}. In thermal equilibrium, this equation adopts the form \cite{Belzig1999}
\begin{equation}
\label{eqn:Eilenberger}
-\bm{v}_{\rm F} \bm{\partial} \hat{g}(\bm{r},\bm{v}_{\rm F},E) = 
\left[ -iE\hat{\tau}_3 + \hat{\Delta}(\bm{r}), \hat{g}(\bm{r},\bm{v}_{\rm F},E) \right],
\end{equation}
where $\hat{g}$ is the quasiclassical Green's function that contains the full information about the
equilibrium properties of the system. This function depends on the energy $E$, the position $\bm{r}$,
the Fermi velocity $\bm{v}_{\rm F}$, and it has the following matrix structure in Nambu (electron-hole) 
space \cite{Belzig1999}
\begin{equation}
\label{eqn:g}
\hat{g} = \bigg( {\begin{array}{cc}
	g & f \\
	f^\dagger & -g \\
\end{array} } \bigg) .
\end{equation}
Moreover, $\hat{\tau}_3 = \mbox{diag}(1,-1)$ is the third Pauli matrix and $\hat{\Delta}$ is the gap 
matrix that contains the information about the modulus and phase of the superconducting order parameter:
\begin{equation}
\label{egn:gap}
\hat{\Delta}(\bm{r}) = \Delta(\bm{r})
\bigg( {\begin{array}{cc}
	0 & e^{i\phi(\bm{r})} \\
	e^{-i\phi(\bm{r})} & 0 \\
	\end{array} } \bigg) = \Delta(\bm{r})\hat{\tau}_{\phi(\bm{r})}.
\end{equation}
Let us also say that the Green's function in Eq.~(\ref{eqn:g}) obeys the normalization condition 
$\hat{g}^2 =\hat{1} \Rightarrow g^2 + ff^\dagger = 1$. On the other hand, in what follows
we shall make use of two additional Pauli matrices: $\hat{\tau}_{\phi}$ from the gap matrix, see
Eq.~(\ref{egn:gap}), and $\hb{\tau}_\phi = i\hat{\tau}_\phi\hat{\tau}_3 = \hat{\tau}_{\phi-\pi/2}$. 
The Pauli matrices introduced in this way obey the standard spin algebra $[\hat{\tau}_3,\hat{\tau}_\phi] 
= 2i\hb{\tau}_\phi$ and the cyclic permutations, the anti-commutation relations 
$\{\hat{\tau}_3,\hat{\tau}_\phi\} = \{\hat{\tau}_3,\hb{\tau}_\phi\} = \{\hat{\tau}_\phi,\hb{\tau}_\phi\} = 0$,
and the normalization conditions $\hat{\tau}_3^2=\hat{\tau}_\phi^2=\hb{\tau}_\phi^2=\hat{1}$.

From now on, our technical task is to solve the Eilenberger equation, see Eq.~(\ref{eqn:Eilenberger}), 
with the appropriate boundary conditions (see below). Once this is done, the knowledge of the quasiclassical 
Green's function allows us to compute any equilibrium property of our system of interest. Thus, for
instance, the local DOS is given by \cite{Belzig1999}
\begin{equation}
\label{eqn:LDOS}
N(\bm{r},\bm{v}_{\rm F},E) = N_0 \Re[{g(\bm{r},\bm{v_{\rm F}},E+i\eta)}],
\end{equation}
where $\eta$ is the broadening parameter and $N_0 =m/\pi$ is the density of states per unit area of a 2D normal metal 
	at the Fermi energy. The Eilenberger equation (\ref{eqn:Eilenberger}) contains the directional derivative 
along the Fermi velocity, which makes this equation effectively one dimensional. This implies that the 
Eq.~(\ref{eqn:LDOS}) gives us the resolved local DOS for a single trajectory of certain length. In order 
to obtain the LDOS in 2D, we need to average Eq.~(\ref{eqn:LDOS}) over all possible trajectories:
\begin{equation}
\label{eqn:ADOS}
N_{\rm 2D}(\bm{r},E) = \langle N(\bm{r},\bm{v}_{\rm F},E) \rangle_{\bm{v}_{\rm F}},
\end{equation}
where $\langle \cdots \rangle_{\bm{v}_{\rm F}}$ stands for the average over the Fermi velocity directions.

Another property of interest in this work is the supercurrent, i.e., the equilibrium current that can
flow through the heterostructure when there is a phase difference between the superconducting electrodes.
The supercurrent density at a temperature $T$ can be expressed in terms of the quasiclassical Green's
functions as follows \cite{Kopnin2001}
\begin{equation}
\label{eqn:j}
\bm{j}(\bm{r}) = -eN_0 \int_{-\infty}^{\infty} \langle \bm{v_{\rm F}} {g}(\bm{r},\bm{v}_{\rm F},E) 
\rangle_{\bm{v}_{\rm F}} \tanh \left( \frac{E}{2T} \right) dE,
\end{equation}
where $e$ is the elementary charge.

\subsection{A fully transparent junction}

\begin{figure}[b!]
\includegraphics[width=0.9\columnwidth,clip]{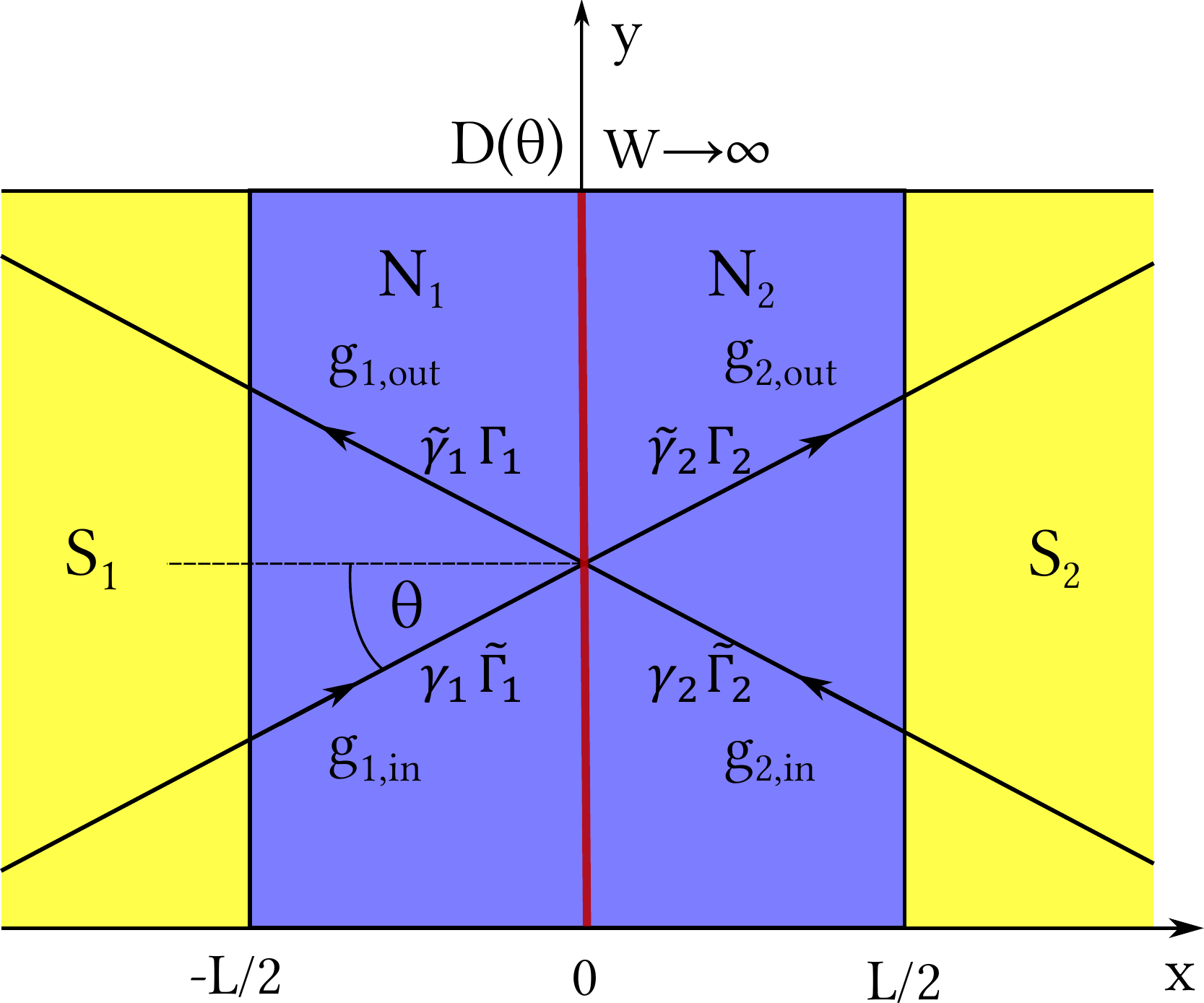}
\caption{(color online) A clean 2D SNS junction with a barrier (red) of transparency $D$ in the middle of 
the normal metal of a length $L$ and infinite width (blue). The coherent functions $\gamma_{1,2}$ and 
$\tilde{\gamma}_{1,2}$ are solutions of the Riccati-like Eilenberger equations, see Eqs.~(\ref{eqn:Riccati1})
and (\ref{eqn:Riccati2}). The functions $\Gamma_{1(2)}$ ($\tilde{\Gamma}_{1(2)}$) are stable solutions 
obtained by integrating the transport equation towards the barrier. The fully transparent case corresponds 
to $D=1$.}
\label{fig-Riccati}
\end{figure}

We first consider a fully transparent (no potential barriers) clean 2D SNS junction of infinite width. 
We assume that the Fermi velocity is along $x$-direction, $\bm{v}_F=v_F\bm{e}_x$ (see Fig.~\ref{fig-Riccati}).
Rewriting Eq.~(\ref{eqn:Eilenberger}) using the Pauli matrix set  $\{\hat{\tau}_\phi, \hb{\tau}_\phi, \hat{\tau}_3\}$ 
allows us to obtain the following set of particular solutions for a spatially inhomogeneous superconductor 
\begin{align}
\hat{g}^s_h(\phi) = &\frac{1}{\Omega}(-iE\hat{\tau}_3 + \Delta\hat{\tau}_\phi), 
\label{egn:gsh}
\\
\hat{g}^s_\pm(\phi,x) = &\frac{1}{2\Omega}(\Delta\hat{\tau}_3 + 
iE\hat{\tau}_\phi \pm i\Omega\hat{\bar{\tau}}_\phi)e^{\pm 2\Omega x/v_{\rm F}} \nonumber \\
 = &\hat{g}_\pm(\phi) e^{\pm 2\Omega x/v_{\rm F}}.
\label{eqn:gspm}
\end{align}
where $\Omega=\sqrt{\Delta^2-E^2}$. Here, $\hat{g}_h^s(x)$ corresponds to an homogeneous solution, while 
$\hat{g}_\pm^s(x)$ are spatially-dependent ones. The general solution of Eq.~(\ref{eqn:Eilenberger}) is 
a linear combination of those and depends on the boundary conditions. For the normal metal we obtain 
correspondingly
\begin{align}
\hat{g}^n_h =&\ \hat{\tau}_3,
\label{gnh}
\\
\hat{g}^n_\pm(x) =&
\hat{\tau}_\pm e^{\pm 2iE x/v_{\rm F}},
\label{gnpm}
\end{align}
where we defined $\hat{\tau}_\pm=\frac{1}{2}(\hat{\tau}_1\pm i\hat{\tau}_2)$. 

We now solve the Eilenberger equation assuming that the order parameter follows a step function:
$\Delta(x) = \Delta\theta(|x| - L/2)$, i.e., there is no inverse proximity effect. For this purpose,
we make following Ansatz for a trajectory starting at $x=-\infty$ in superconductor 1 with the phase 
$-\varphi/2$ going straight through the normal metal with a length $L$ and ending in $x=\infty$ in 
superconductor 2 with the phase $+\varphi/2$:
\begin{align}
\hat{g}^s\left(x<-L/2\right)= &\hat{g}^s_h(-\varphi/2)+ B_1\hat{g}_+(-\varphi/2)e^{2\Omega x/v_{\rm F}}
\\
\hat{g}^s\left(x>L/2\right)= &\hat{g}^s_h(\varphi/2) + B_2\hat{g}_-(\varphi/2)e^{-2\Omega x/v_{\rm F}}
\\
\hat{g}^n\left(x\right)= &A\hat{g}^n_h+ A_-\hat{g}^n_-(x)+ A_+\hat{g}^n_+(x),
\end{align}
where $A, A_\pm, B_{1,2}$ are unknown coefficients, which have to be determined with the help of the
boundary conditions at the interfaces. We assume that the Green's function is a continuous function 
throughout the system, which leads to the boundary conditions at the two SN interfaces
\begin{equation}
\label{eqn:boundary_conditions}
\begin{split}
\hat{g}^s(-L/2)&=\hat{g}^n(-L/2),
\\
\hat{g}^n(L/2)&=\hat{g}^s(L/2).
\end{split}
\end{equation}
Using these boundary conditions and solving the problem in the opposite direction 
($\varphi/2 \rightarrow -\varphi/2$), we arrive at the following solution for the Green's function 
inside the normal metal
\begin{widetext}
\begin{equation}
\label{eqn:g_solution}
g^n(\varphi,\sigma,E,\theta) = A = -i\frac{E + \sqrt{\Delta^2 -E^2}\tan(E L/v_{\rm F}\cos(\theta) 
+\sigma \varphi/2)}{\sqrt{\Delta^2 - E^2} - 
E\tan(E L/v_{\rm F}\cos(\theta)  +\sigma\varphi/2)},
\end{equation}
\end{widetext}
where $\sigma=\pm$ denotes the direction of propagation (left of right), $v_{\rm F} $ is the magnitude of the Fermi velocity and $\theta$ is the angle between the incoming trajectory and the direction perpendicular to the interface (see Fig.~\ref{fig-Riccati}). We note that $g^n(\varphi,\sigma,E,\theta)$ does not depend on the position, 
i.e., it is constant throughout the normal  metal. The LDOS can now be obtained from Eq.~(\ref{eqn:LDOS})
\begin{equation}
\label{eqn:LDOS_single_trajectory}
N(\varphi,\sigma,E,\theta=0) = N_0\Re[g^n(\varphi,\sigma, E+i\eta,\theta=0)],
\end{equation}
which gives us the resolved LDOS for a single trajectory of a length $L$.
The LDOS in 2D, see Eq.~(\ref{eqn:ADOS}), adopts in this case the form
\begin{equation}
\label{ADOS_integral}
N_{\rm 2D}(\varphi,E) = sum_{\sigma=\pm 1}\frac{1}{\pi}\int_{-\pi/2}^{\pi/2}
N(\varphi,\sigma,E,\theta) d\theta.
\end{equation}

\subsection{Effect of a finite transparency}

To investigate the role of a finite transparency through the heterostructure, we consider now
a SNS junction with a normal metal of length $L$ and infinite width featuring a potential
barrier in the middle ($x=0$), see Fig.~\ref{fig-Riccati}. The barrier is characterized by the transmission 
coefficient $D$ and the corresponding reflection coefficient is denoted by $R$ ($R=1-D$). The
angular dependence of the $D$ is taking from a delta-like potential and it is given by \cite{Eschrig2000}
\begin{equation}
\label{eqn:D}
D(\theta) = \frac{D_0\cos^2{\theta}}{R_0 + D_0\cos^2{\theta}},
\end{equation}
where $D_0$ is the transmission coefficient for $\theta =0$, i.e., for the trajectory perpendicular 
to the interface and $R_0 = 1 - D_0$.

In order to solve the problem analytically it is convenient to use the so-called Riccati 
parametrization in which for the quasiclassical Green's function is parametrized in terms
of two coherent functions as follows \cite{Eschrig2000}
\begin{equation}
\label{eqn:Riccati}
\hat{g}(\bm{r},\bm{v}_F,E) = \frac{1}{1+\gamma\tilde{\gamma}}\bigg( {\begin{array}{cc}
	1-\gamma\tilde{\gamma} &2\gamma \\
	2\tilde{\gamma} & -1+\gamma\tilde{\gamma} \\
	\end{array} }\bigg).
\end{equation}
With this parametrization the normalization condition $\hat{g}^2 = \hat{1}$ is automatically
fulfilled and from the Eilenberger equation, see Ref.~(\ref{eqn:Eilenberger}), one can show
that the coherent functions $\gamma$ and $\tilde \gamma$ satisfy the following decoupled first-order
differential equations \cite{Eschrig2000}
\begin{align}
-\bm{v}_F\bm{\partial}\gamma(\bm{r}) = -2iE\gamma(\bm{r}) + \Delta^*\gamma(\bm{r})^2 - \Delta(\bm{r}),
\label{eqn:Riccati1}
\\
\bm{v}_F\bm{\partial}\tilde{\gamma}(\bm{r}) = -2iE\tilde{\gamma}(\bm{r}) + 
\Delta\tilde{\gamma}(\bm{r})^2 - \Delta^*(\bm{r}).
\label{eqn:Riccati2}
\end{align} 

We now follow Ref.~\cite{Eschrig2000} and define the coherent functions on the both 
sides of the barrier $\gamma_1,\tilde{\gamma}_1,\gamma_2,\tilde{\gamma}_2$, which are the stable 
solutions for integration towards the interface. The boundary conditions determine the solutions 
away from the interface denoted by $\Gamma_1,\tilde{\Gamma}_1,\Gamma_2,\tilde{\Gamma}_2$ (see 
Fig.~\ref{fig-Riccati})
 \begin{align}
\Gamma_{1,2}=R_{1,2}\gamma_{1,2}(0)+D_{1,2}\gamma_{2,1}(0),
\label{eqn:Gamma}
\\
\tilde{\Gamma}_{1,2}=\tilde{R}_{1,2}\tilde{\gamma}_{1,2}(0)+\tilde{D}_{1,2}\tilde{\gamma}_{2,1}(0),
\label{eqn:Gamma_tilde}
\end{align}
where $R_1,D_1$ and $\tilde{R}_1,\tilde{D}_1$ are given by
\begin{align}
R_1=R\frac{1+\gamma_2\tilde{\gamma}_2}{1+R\gamma_2\tilde{\gamma}_2+D\gamma_1\tilde{\gamma}_2},
\\
D_1=D\frac{1+\tilde{\gamma}_2\gamma_1}{1+R\gamma_2\tilde{\gamma}_2+D\gamma_1\tilde{\gamma}_2},
\\
\tilde{R}_1=R\frac{1+\gamma_2\tilde{\gamma}_2}{1+R\gamma_2\tilde{\gamma}_2+D\gamma_2\tilde{\gamma}_1},
\\
\tilde{D}_1=D\frac{1+\gamma_2\tilde{\gamma}_1}{1+R\gamma_2\tilde{\gamma}_2+D\gamma_2\tilde{\gamma}_1}.
\end{align}
The coefficients $R_2,D_2$ and $\tilde{R}_2,\tilde{D}_2$ are given by the analogous expressions. All 
the previous expressions fulfill $R_j+D_j = 1$ and $\tilde{R}_j+\tilde{D}_j=1$. 

To show how to obtain the expression for the quasiclassical Green's function, we consider here the
solution for $\gamma_1$ (Fig.~\ref{fig-Riccati}). The solution for $\gamma_1$ of Eq.~(\ref{eqn:Riccati1}) 
in a homogeneous superconductor (with the superconducting phase $\phi=\varphi/2$) and a normal metal 
are respectively
\begin{eqnarray}
\gamma_1^s(\varphi,E) & = & \frac{\Delta e^{i\varphi/2}}{\sqrt{\Delta^2-E^2} - iE}, \\
\gamma_1^n(\varphi,\sigma,E,\theta) & = & A e^{2iEx/\sigma v_F\cos(\theta)},
\end{eqnarray}
where $A$ is an unknown coefficient. By applying the boundary condition of Ref.~(\ref{eqn:boundary_conditions}) 
at the left SN interface $(x=-L/2)$ we obtain the solution in the normal metal as
\begin{equation}
\gamma^n_1(\varphi,\sigma,E,\theta) =\frac{\Delta}{\sqrt{\Delta^2-E^2} - iE}e^{i\varphi/2 + iEL/\sigma v_F\cos(\theta)}.
\label{eqn:gamma1}
\end{equation}
By repeating the same procedure for the other $\gamma$ ($\tilde{\gamma}$) functions we obtain the 
full set of  solutions in the normal metal
\begin{align}
\tilde{\gamma}^n_1(\varphi,\sigma,E,\theta) = & 
\frac{\Delta}{\sqrt{\Delta^2-E^2} - iE}e^{-i\varphi/2 + iEL/\sigma v_F\cos(\theta)},
\\
\gamma^n_2(\varphi,\sigma,E,\theta) = & 
\frac{\Delta}{\sqrt{\Delta^2-E^2} - iE}e^{-i\varphi/2 + iEL/\sigma v_F\cos(\theta)},
\\
\tilde{\gamma}^n_2(\varphi,\sigma,E,\theta) = & 
\frac{\Delta}{\sqrt{\Delta^2-E^2} - iE}e^{i\varphi/2 + iEL/\sigma v_F\cos(\theta)}.
\label{eqn:gamma_tilde_2}
\end{align}
Notice that since the LDOS does not depend on the position, we have omitted the spatial arguments in 
the coherent functions in Eqs.~(\ref{eqn:gamma1})-(\ref{eqn:gamma_tilde_2}). 

Now, the $g$ component of the incoming Green's function ($g_{1, \rm in} $ in the Fig.~\ref{fig-Riccati}) 
is obtained from \cite{Eschrig2000}
\begin{equation}
g_{1, \rm in} = \frac{1-\gamma_1^n\tilde{\Gamma}_1^n}{1+\gamma_1^n\tilde{\Gamma}_1^n},
\end{equation}
where $\tilde{\Gamma}_1^n$ is defined in Eq.~(\ref{eqn:Gamma_tilde}). Analogously, one can define 
the outgoing Green's function ($g_{1, \rm out} $ in the Fig.~\ref{fig-Riccati}) arriving at the solutions
\begin{align}
g_{1,\rm in} = \frac{(1-\gamma^n_1\tilde{\gamma}^n_1)(1+\gamma^n_2\tilde{\gamma}^n_2) + 
D(\gamma^n_2+\gamma^n_1)(\tilde{\gamma}^n_1-\tilde{\gamma}^n_2)}
{(1+\gamma^n_1\tilde{\gamma}^n_1)(1+\gamma^n_2\tilde{\gamma}^n_2) + D(\gamma^n_2-\gamma^n_1)
(\tilde{\gamma}^n_1-\tilde{\gamma}^n_2)},
\label{eqn:g1in}
\\
g_{1,\rm out} = \frac{(1-\gamma^n_1\tilde{\gamma}^n_1)(1+\gamma^n_2\tilde{\gamma}^n_2) + 
D(\gamma^n_1-\gamma^n_2)(\tilde{\gamma}^n_1+\tilde{\gamma}^n_2)}
{(1+\gamma^n_1\tilde{\gamma}^n_1)(1+\gamma^n_2\tilde{\gamma}^n_2) + 
D(\gamma^n_2-\gamma^n_1)(\tilde{\gamma}^n_1-\tilde{\gamma}^n_2)}.
\label{eqn:g1out}
\end{align}
Finally, the total Green's function in the normal metal is the average of the incoming and the outgoing 
ones
\begin{equation}
\label{eqn:g_barrier}
g^n_1(\varphi,\sigma,E,\theta) = \frac{1}{2}\big[g_{1,\rm in}(\varphi,\sigma,E,\theta) +
g_{1,\rm out}(\varphi,\sigma,E,\theta)\big].
\end{equation}
The single trajectory and the  2D LDOS are obtained by inserting Eq.~(\ref{eqn:g_barrier}) in 
Eqs.~(\ref{eqn:LDOS}) and (\ref{eqn:ADOS}), respectively. 

\subsection{Presence of a weak magnetic field and the effect of the finite width} \label{MField}

\begin{figure}[t]
\includegraphics[width=0.9\columnwidth,clip]{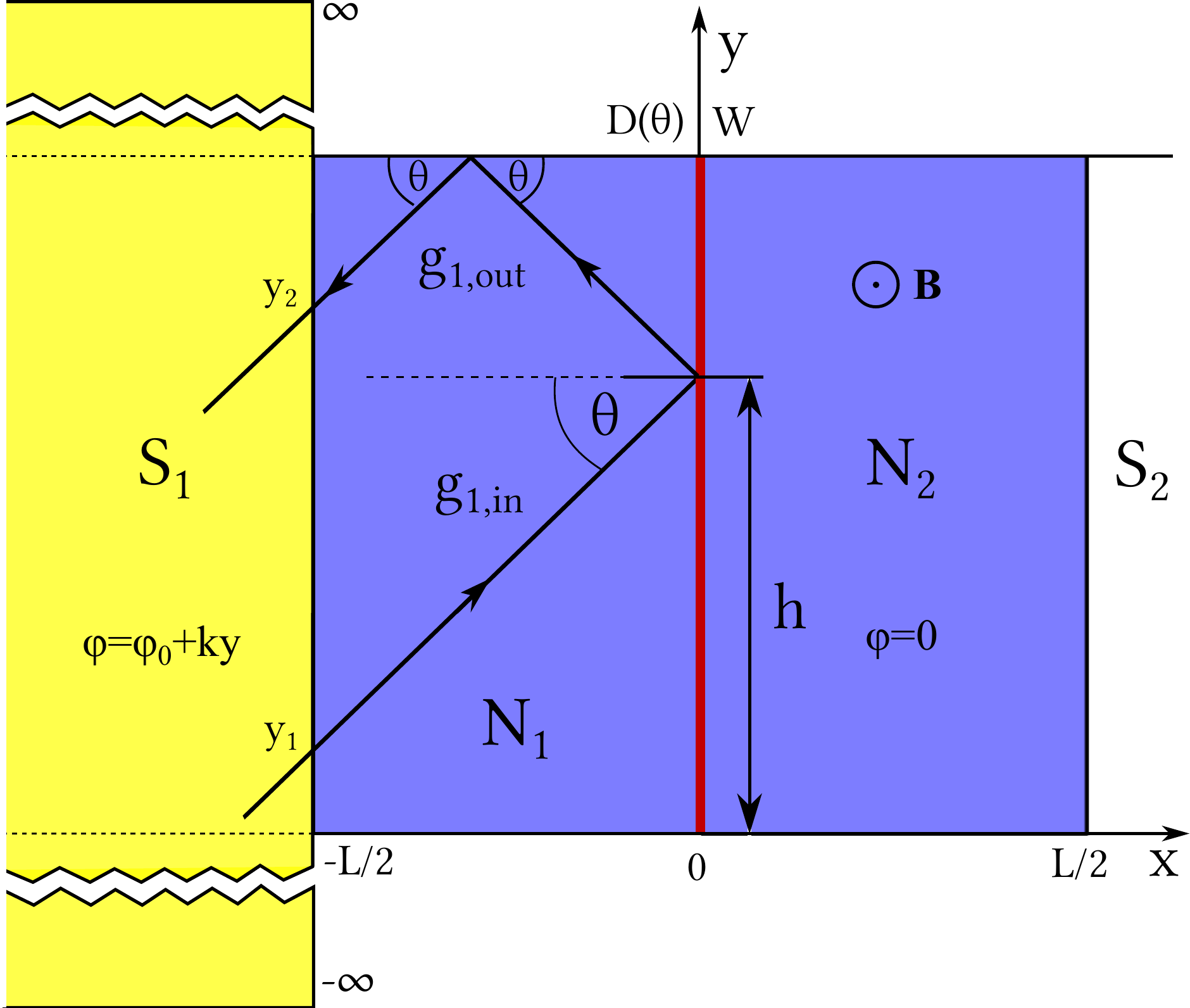}
\caption{(color online) A clean 2D SNS junction in the presence of a weak magnetic field. 
The normal region contains a potential with transparency $D(\theta)$ and it has a length $L$
and a finite width $W$. The quasiparticles are assumed to undergo specular reflection at 
the interface between the normal metal and the vacuum. The superconducting phase difference 
is a linear function of $y$ due to the magnetic field.}
\label{fig-Bfield}
\end{figure}

We now want to describe the effect of the presence of a weak (perpendicular) magnetic field 
and also consider the effect of having a finite width $W$ in the normal region. By weak 
magnetic field we mean that one can neglect the orbital and Zeeman effects in the normal
region and the role of the field is simply to spatially modulate the superconducting
phase inside the electrodes. In other words, the magnetic field only enters via the
gauge-invariant superconducting phase difference that becomes \cite{Tinkham1996} 
\begin{equation}
\label{egn:phi_Mfield}
\varphi(y) = \varphi_0 + 2\pi \left( \frac{\Phi}{\Phi_0} \right) \frac{y}{W},
\end{equation} 
where $\varphi_0$ is a constant value superconducting phase difference, $\Phi$ is the magnetic
flux enclosed in the normal region, $\Phi_0=h/(2e)$ is the flux quantum, and $y$ is the transverse
coordinate (parallel to the SN interfaces), see Fig.~\ref{fig-Bfield}.

We assume that in the normal region the quasiparticles are specularly reflected in the interfaces
between the normal metal and the vacuum. Moreover, for the sake of simplicity, we consider only 
processes with one specular reflection. With this assumption, the range for $\theta$, the angle
defining the quasiparticle trajectory, depends on the geometrical parameters of the junction as
follows: $-\theta_0(h) < \theta < \theta_0(h)$, where $\theta_0(h) = \arctan[2d(h)/L]$, and $d(h)=h$ for 
$h \leq W/2$ and $d(h)=W-h$ for $h>W/2$ (see Fig.~\ref{fig-Bfield} for a definition of $h$). 

To avoid reflections inside the superconductors we assume they are infinitely wide. From 
Eq.~(\ref{egn:phi_Mfield}) we can see that the superconducting phase difference depends on the 
$y$ position of the incoming and outgoing Green's function at the SN interface $(x=-L/2)$. 
The averaged LDOS over all trajectories, see Eq.~(\ref{eqn:ADOS}), adopts the form
\begin{align}
	 & N_{\rm 2D}(\varphi,E) = \label{eqn:ADOS_Mfield}\\ \nonumber	
	 & \sum_{\sigma=\pm 1} \frac{1}{W}\int_0^W\frac{1}{2\theta_0(h)} 
		\int_{-\theta_0(h)}^{\theta_0(h)} N(\varphi,\sigma,E,\theta,h) d\theta dh.
\end{align}
%
 
\subsection{The Josephson current} \label{sec-supercurrent}

One of the questions that we discuss below is the relation between the LDOS and the Josephson
current that flows across the junction. To establish such a relation, it is convenient to 
consider the case of a single trajectory at normal incidence. In this case, the Josephson current, see Eq.~(\ref{eqn:j}),
can be expressed as
\begin{equation}
\label{eqn:j_singleTrajecotry}
I(\varphi) = -eN_0 v_{\rm F} W \int_{-\infty}^{\infty} J_s(\varphi,E,\theta=0) 
\tanh \bigg( \frac{E}{2T} \bigg) dE ,
\end{equation}
where $W$ is the width of the normal metal and $J_s(\varphi,E,\theta) = (1/2)\Re
[g^n(\varphi,E,1,\theta) - g^n(\varphi,E,-1,\theta)]$ is the spectral current.
By inserting Eq.~(\ref{eqn:g_solution}) in the formula above, one obtains the single-trajectory 
spectral current
\begin{widetext}
	\begin{equation}
	\begin{split}
	\label{eqn:js}
	J_s(\varphi,E) & =\Im\left\{ \frac{1}{4}\frac{\Delta^2\sin\varphi}
	{(\Delta^2-2(E+i\eta)^2) \cos(2(E+i\eta)L/v_{\rm F}) - 
	2(E+i\eta) \sqrt{\Delta^2-(E+i\eta)^2} \sin(2(E+i\eta)L/v_{\rm F}) + 
\Delta^2\cos\varphi}\right\}
	\\
	&=-\Re\left\{\sum_{\sigma=\pm}\int_0^E \bigg( \frac{v_{\rm F}} {\sqrt{\Delta^2-(\epsilon+i\eta)^2}} +L \bigg)
	\frac{ \partial g^n(\varphi,\sigma,\epsilon+i\eta,\theta=0)}{\partial \varphi} d\epsilon,\right\}
	\end{split}
	\end{equation}
\end{widetext}
where $g^n(\varphi,\sigma,\epsilon,\theta)$ is given by Eq.~(\ref{eqn:g_solution}). At zero temperature $\tanh(E/2T) \rightarrow \sign(E)$. 
\section{Results} \label{results}

In this section we shall make use of the formalism developed in the previous one and explore 
systematically the role of different parameters in the LDOS of a clean SNS junction such as 
the length, the width, the transmission barrier, and the presence of a weak external magnetic 
field.

\subsection{The single trajectory case}

\begin{figure}[t]
\includegraphics[width=0.9\columnwidth,clip]{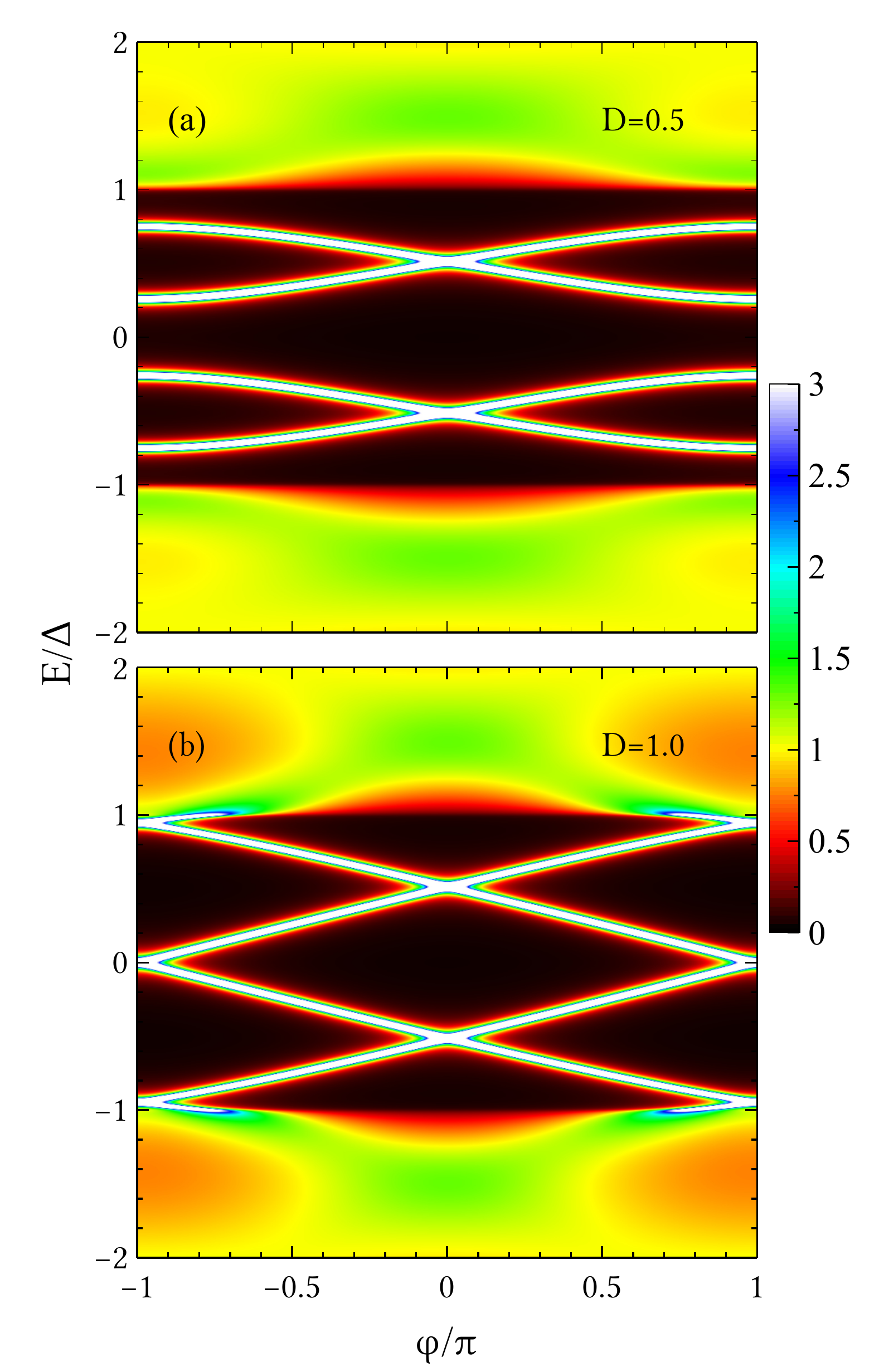}
\caption{(color online) The single-trajectory LDOS for a SNS junction as a function of energy $E$ and 
superconducting phase difference $\varphi$ for transparencies (a) $D=0.5$ and (b) $D=1.0$. The length of 
the trajectory in both panels is $L=2.0 \xi$ and the broadening parameter was taken $\eta=0.01\Delta$.}
\label{fig-LDOS-single}
\end{figure}

For illustration purposes, we start in this subsection with the analysis of the trajectory-resolved 
LDOS in the absence of a magnetic field. In Fig.~\ref{fig-LDOS-single} we present the LDOS inside the
normal region for a single trajectory of length $L=2.0\xi$ as a function of both the energy and the 
superconducting phase different for two values of the barrier transmission, $D=0.5$ and $D=1.0$. 
Let us recall that the LDOS is constant throughout the normal metal. As expected, the main feature
of the LDOS is the presence of Andreev bound states (ABBs) inside the gap that evolve with the 
phase difference. In particular, we see the appearance of four different ABSs for this value of
the length trajectory. Notice that in the fully transparent case, see Fig.~\ref{fig-LDOS-single}(b), 
the ABS energy is basically a linear function of the phase and, in particular, two states have 
zero energy (i.e., they appear at the Fermi level) at $\varphi=\pm\pi$. In contrast, at finite
transparency, see Fig.~\ref{fig-LDOS-single}(a), there is a gap between the ABSs, which we shall
term Andreev gap, irrespective of the value of the phase difference. In the case of perfect
transparency, and with the help of Eqs.~(\ref{eqn:g_solution}) and (\ref{eqn:LDOS_single_trajectory}),
one can show that the energies of the ABSs for a single trajectory are given by the solutions of 
the following well-known equation \cite{Kulik1970}
\begin{equation}
\label{Andreev_bound_states}
\frac{2EL}{v_{\rm F}} \pm \varphi - 2\arccos \left( \frac{E}{\Delta} \right) = 2\pi n,
\end{equation} 
where $n$ is an integer number, $v_{\rm F}$ is the Fermi velocity, and $\varphi$ is the superconducting
phase difference difference. For long trajectories ($L \gg \xi=v_{\rm F}/\Delta$) the previous equation
reduces to (for energies much smaller than $\Delta$)
\begin{equation}
\label{Andreev_bound_states_long}
\frac{2EL}{v_{\rm F}} \pm \varphi = (2n+1) \pi .
\end{equation} 
From this expression we see that in this long-junction limit the energy of the ABSs depends linearly on 
the phase, something that it is already apparent in Fig.~\ref{fig-LDOS-single}(b).

\subsection{The 2D case}  

\begin{figure}[t]
\includegraphics[width=0.9\columnwidth,clip]{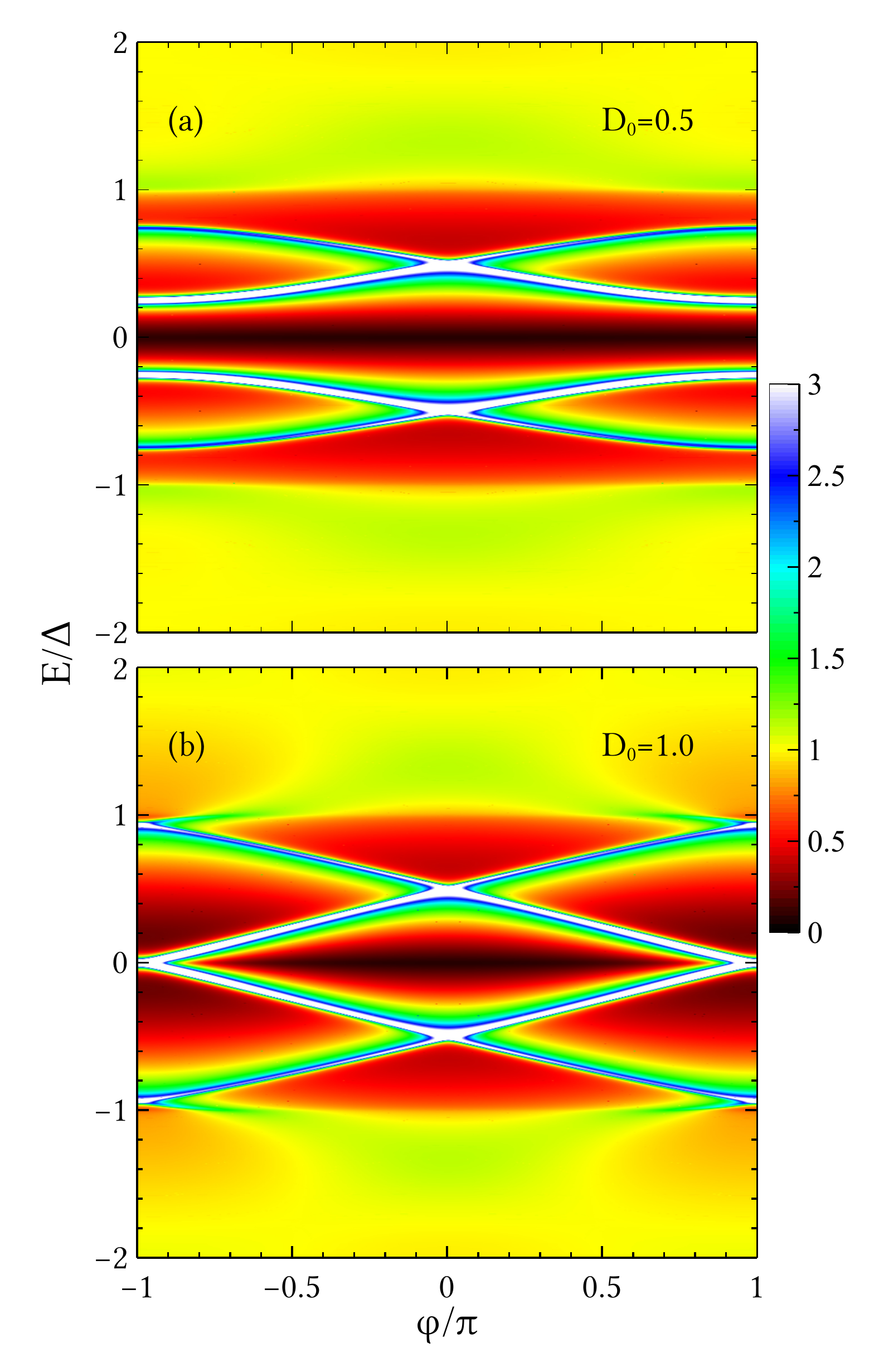}
\caption{(color online) Averaged LDOS of a SNS junction of length $L=2.0\xi$ as a function of energy $E$ and 
superconducting phase difference $\varphi$ for normal incidence transparencies (a) $D_0=0.5$ and (b) $D_0=1.0$. 
The broadening parameter was taken $\eta=0.01\Delta$.}
\label{fig-LDOS-averaged}
\end{figure}

Let us turn now to the analysis of the angle-averaged LDOS in the normal metal for junctions of 
infinite width and in the absence of an external magnetic field. In Fig.~\ref{fig-LDOS-averaged} we 
present the LDOS inside the normal region for a junction of length $L=2.0\xi$ as a function of both the 
energy and the phase difference. The two panels correspond to two different values of the barrier
transmission for normal incidence, $D_0=0.5$ and $D_0=1.0$. Let us recall that we are using an
angular dependence of the transmission coefficient given by Eq.~(\ref{eqn:D}). As one can observe,
this angle-averaged LDOS exhibits many of the feature of the single-trajectory case, see 
Fig.~\ref{fig-LDOS-single}, the main difference being the larger DOS inside the gap due to the
contributions of trajectories of different lengths. In particular, we still see that the role of
the finite transparency is to induce a finite and hard Andreev gap for any phase value, while such 
a gap vanishes in the case $D_0=1.0$ for $\varphi=\pm\pi$. 

\begin{figure*}[t]
\includegraphics[width=\textwidth,clip]{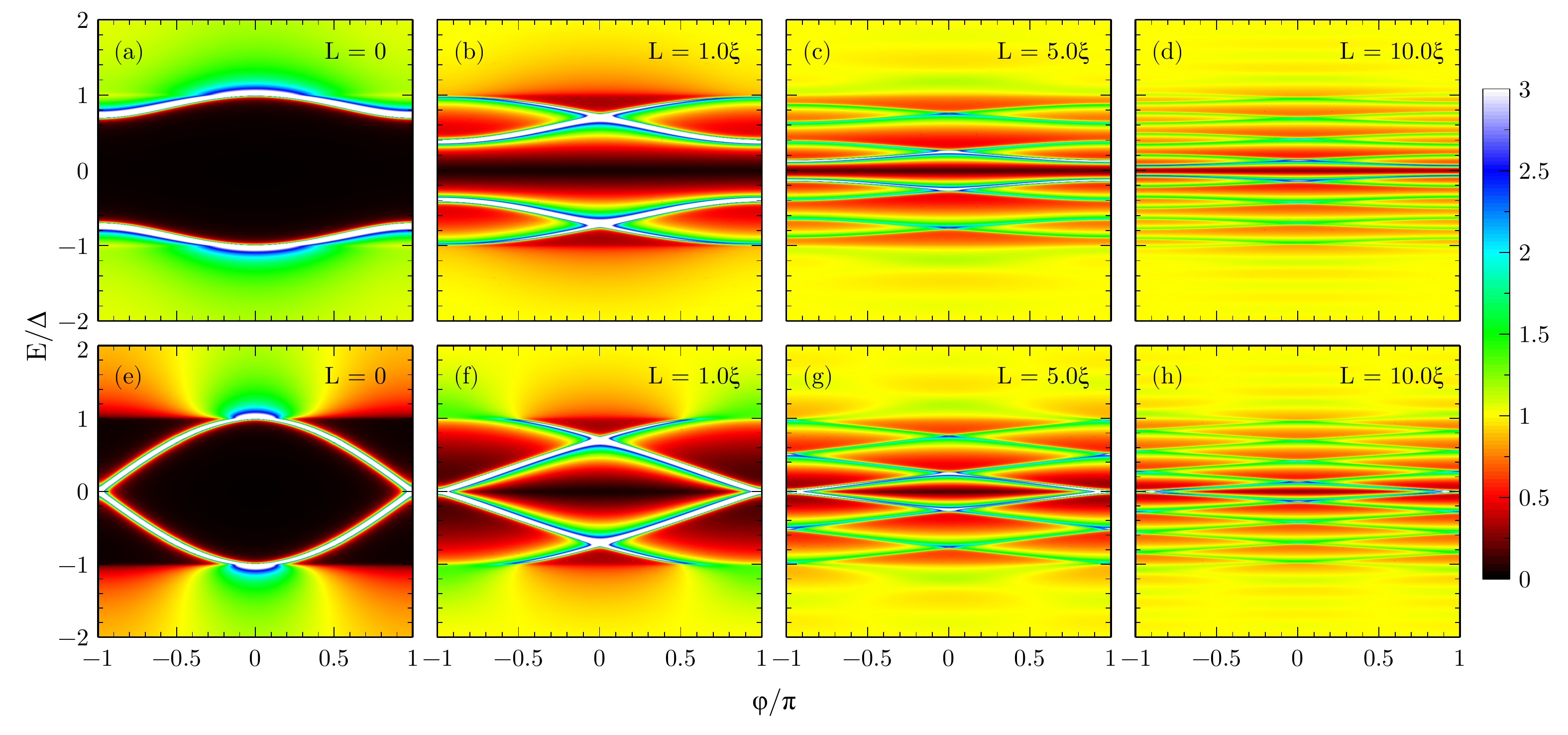}
\caption{(color online) The averaged LDOS in the normal region of a SNS junction as a function of 
energy and the superconducting phase difference. The different panels correspond to different values
of the junction length, as indicated in the legends, and to two different values of the barrier
transparency: (a)-(d) $D_0=0.5$ and (e)-(h) $D_0=1.0$. The width is considered to be infinite and 
the broadening parameter is $\eta=0.01\Delta$.}
\label{fig-LDOS-length}
\end{figure*}

To understand the role of the junction length, we present in Fig.~\ref{fig-LDOS-length} the results 
for the LDOS by varying the length from the short-junction case ($L \ll \xi$) to the long-junction
limit ($L \gg \xi$). As one can see, the number of the ABSs increases with increasing junction 
length. In the short-junction limit ($L\rightarrow 0$), see panels (a) and (e), the LDOS exhibits 
the same behavior as in the single-trajectory case due to absence of the contributions of trajectories 
of various lengths. The Andreev spectrum of a junction with the intermediate normal metal length  
($L=1.0\xi$, see Fig.~\ref{fig-LDOS-length}(b,f)) is similar to the one shown before in 
Fig.~\ref{fig-LDOS-averaged}. By increasing the junction length the Andreev gap diminishes,
see Fig.~\ref{fig-LDOS-length}(c,g) for $L=5\xi$, and the proximity effect tends to disappear
altogether for very long junctions, see Fig.~\ref{fig-LDOS-length}(c,g) for $L=10\xi$. 

\subsection{Presence of a weak magnetic field and the effect of a finite width} \label{megnetic_field}

In experimental setups, like that of Ref.~\cite{Bretheau2017}, the superconducting 
phase difference is controlled by incorporating the weak link into a superconducting loop and applying
a weak magnetic field. For this reason, we analyze in this section the role of the application of a
weak external magnetic field perpendicular to the SNS junction and study also the role of having a 
finite width. As explained in section \ref{MField}, by weak magnetic field we mean that the only role of the
external field is to modulate the superconducting phase difference inside the electrodes and along the
SN interfaces. This modulation leads to the following expression for the gauge-invariant phase difference
\begin{equation}
\label{eqn:phi_Mfield1}
\varphi(y) = \varphi_0 + 2\pi \left( \frac{\Phi}{\Phi_0} \right) \frac{y}{W},
\end{equation}
where $y$ is the transverse coordinate along the SN interfaces (see Fig.~\ref{fig-Bfield}), $\varphi_0$ 
is a constant, $\Phi$ is the magnetic flux enclosed in the normal region, and $\Phi_0$ is the flux quantum.

\begin{figure*}[t]
\includegraphics[width=\textwidth,clip]{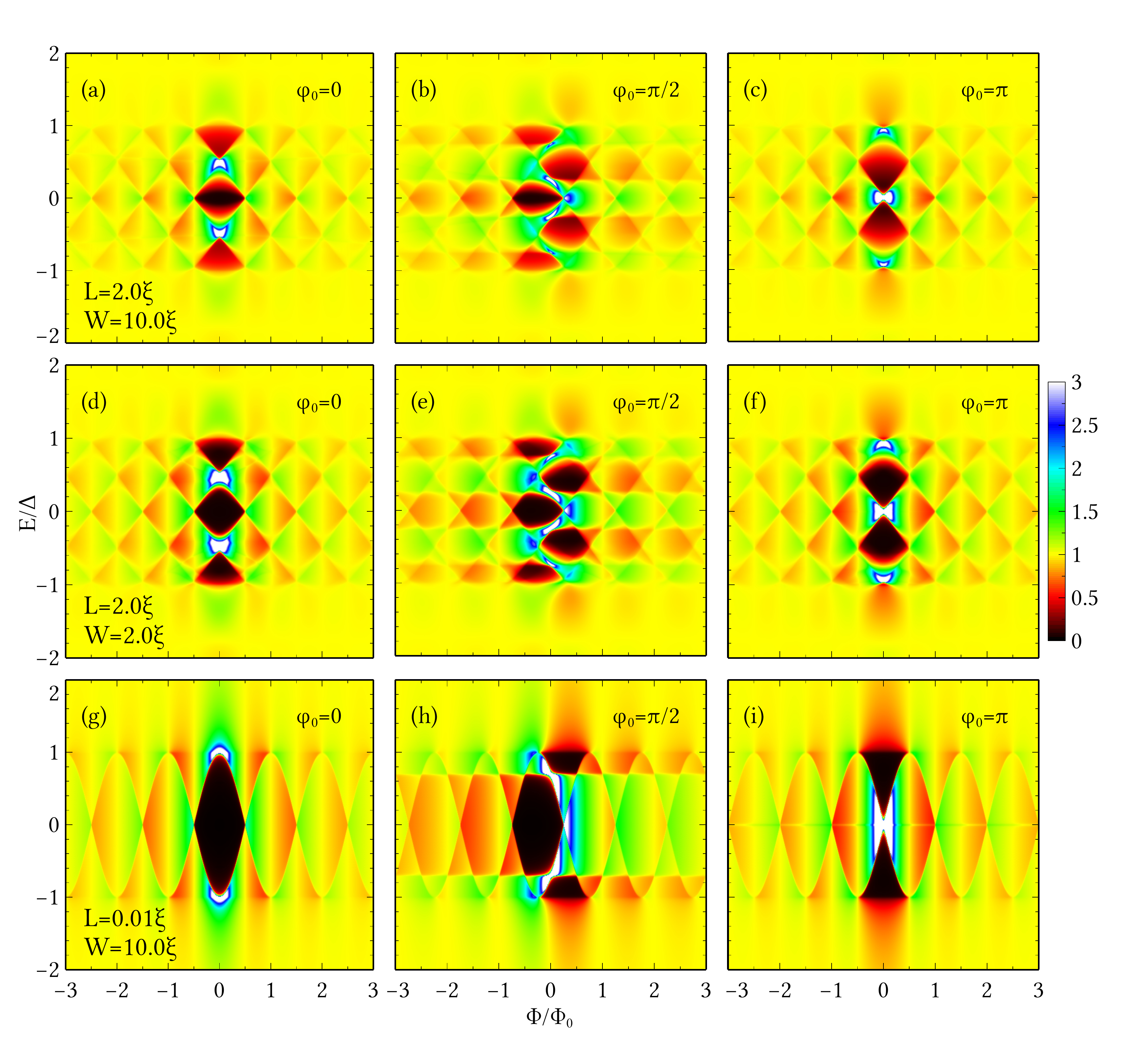}
\caption{(color online) Averaged LDOS in fully transparent 2D SNS junctions as a function of energy 
and the magnetic flux $\Phi$ enclosed in the junction for several values of the constant part of the
superconducting phase difference $\varphi_0$. Panels (a)-(c) correspond to a length $L=2.0\xi$ and
a width $W=10\xi$, (d)-(f) to $L=2.0\xi$ and $W=2.0\xi$, and (g)-(i) to $L=0.01\xi$ and $W=10\xi$. 
The broadening parameter was taken in all cases $\eta=0.01\Delta$.}
\label{fig-LDOS-Bfield}
\end{figure*}

Here, we focus on the case in which the magnetic field is only applied to the junction and the constant
part of the superconducting phase difference, $\varphi_0$, can take an arbitrary value. In the next
section we shall consider the situation where the junction is incorporated into a superconducting 
loop and $\varphi_0$ is determined by the magnetic flux enclosed in the loop. Making use of the 
formalism detailed in section \ref{MField}, we have computed the results shown in Fig.~\ref{fig-LDOS-Bfield} 
for the averaged LDOS as a function of energy and the magnetic flux enclosed in the junction for
different values of the length and width of the normal region and the phase $\varphi_0$. In particular,
Fig.~\ref{fig-LDOS-Bfield}(a-c) show the results for the case of a junction with an intermediate length 
($L=2.0\xi$) and the width $W=10.0\xi$. In this case and for weak magnetic fields ($\Phi\lesssim 0.5\Phi_0$),
the features related to the ABSs are smeared but they are still clearly visible. In the cases of 
$\varphi_0=0,\pi$, the ABSs are visible as peaks centered around the zero magnetic field, while in the 
case $\varphi_0=\pi/2$ the peaks are shifted to $\Phi=\Phi_0/4$. For stronger magnetic fields $\Phi>\Phi_0$,
and irrespective of the value of $\varphi_0$, the features related to the ABSs are strongly suppressed 
due to the destructive interference between different quasiclassical trajectories that see effectively
different values of the phase difference. Notice also that all the structures are symmetric with respect 
to the Fermi energy ($E=0$), but the symmetry with respect to zero magnetic field does not hold in the case 
of $\varphi_0=\pi/2$. 

In Fig.~\ref{fig-LDOS-Bfield}(d-f) we also show the results for an intermediate junction length $L=2.0\xi$,
but this time the junction is narrower with $W=2.0\xi$. Comparing the results with those of the much wider junction shown in Fig.~\ref{fig-LDOS-Bfield}(a-c),
we see that the width of the junction does not have a very strong impact. This can be explained by with the help
of Eq.~(\ref{eqn:phi_Mfield1}). That formula tells us that the range phases $\varphi(y)$
as function of $y$ is independent of the width $W$ and only depends on the magnetic flux: $\varphi \in 
[\varphi_0,\varphi_0 + 2\pi \Phi/\Phi_0]$. Hence, the phase pattern in Fig.~\ref{fig-LDOS-Bfield}(a-c) and (d-f) are practically the same for the equal phase biases $\varphi_0$. From the discussions above, it is obvious that the largest 
contributions to the Andreev spectrum come from the shortest trajectories. The main difference therefor appears in the slightly large Andreev gap for $W=2\xi$ compared to the case $W=10\xi$ due to absence of the contribution of long trajectories. 

To explore the role of the junction in the presence of a weak magnetic field, we present the results for
a junction of length $L=0.01\xi$ and width $W=10\xi$ in Fig.~\ref{fig-LDOS-Bfield}(g-i). As in previous 
cases, for the weak fields ($\Phi\lesssim\Phi_0/2$) the ABS peaks are smeared but still visible, while 
for stronger magnetic fields they dissappear. The peaks for $\varphi_0=0,\pi$ are located around zero 
field, while for $\varphi_0=\pi/2$ they are shifted to $\Phi=\Phi_0/4$. The Andreev gap is empty in this
case because we only have contributions from short trajectories. In the case of $\varphi_0=0$, panel (g),
the Andreev peaks are shifted to the edge of the gap $(E\approx\Delta)$. For $\varphi_0=\pi$, see panel
(i), we observe only one ABS in each branch of the spectrum in contrast to the case of $L=2.0\xi$ where 
we have two. The geometric symmetry of the pattern is similar to the previous cases.

\section{Discussion} \label{discussion}

\subsection{LDOS of a SNS junction embedded in a superconducting loop}
\begin{figure*}[t]
	\includegraphics[width=\textwidth,clip]{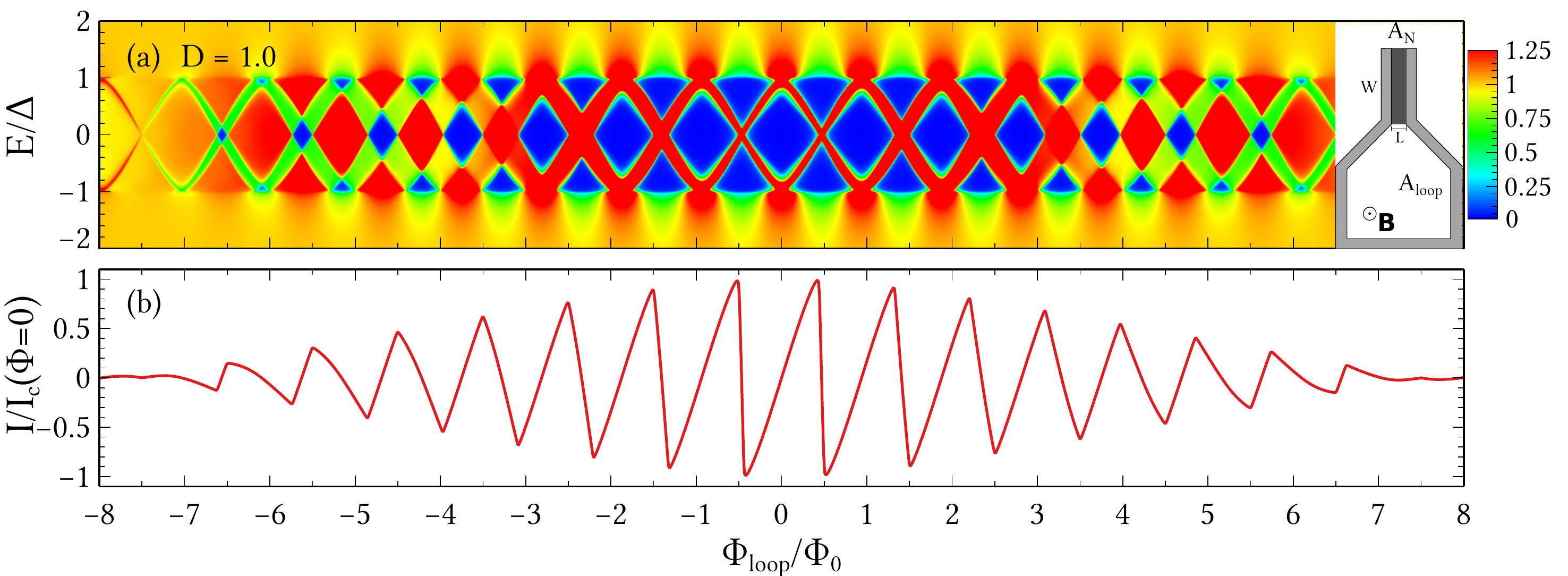}
	\caption{(color online) (a) Averaged LDOS in the normal region of a clean SNS junction embedded 
		in a superconducting loop, as schematically shown in the right inset. The LDOS is shown as
		a function of the energy and the magnetic flux through the entire loop of area $A_{\rm loop}$.
		In this case the SNS junction is fully transparent, the length is $L=0.2\xi$, the width $W=0.7\xi$, 
		and the broadening parameter $\eta=0.01\Delta$. (b) The corresponding Josephson current normalized
		by the critical current at zero field.} 
	\label{fig-LDOS-loop}
\end{figure*}

As mentioned above, the practical way to investigate the phase dependence of the LDOS in a weak 
link is by incorporating it into a superconducting loop and applying an external magnetic field.
This is, for instance, what it was done in Ref.~\cite{Bretheau2017} in which
the authors used graphene as a normal metal in the weak link. Inspired by this experiment, we now 
consider a setup like the one shown in the inset of Fig.~\ref{fig-LDOS-loop}(a) where the SNS junction 
of area $A_{\rm N}$ is embedded in a superconducting loop of total area $A_{\rm loop}$. We assume,
like in the experiments, that an external magnetic field is applied such that the total magnetic
flux enclosed in the whole loop is equal to $\Phi_{\rm loop}$. This flux determines now the constant part of
the phase difference, $\varphi_0$, which is given by $\varphi_0 = 2\pi \Phi_{\rm loop}/\Phi_0$. Thus, the
gauge-invariant phase difference is modulated along the SN interfaces as 
\begin{equation}
\label{eqn:Phi_loop}
	\varphi_{\rm loop}(y) = 2 \pi \frac{\Phi_{\rm loop}}{\Phi_0} 
	\left( 1 + \frac{A_{\rm N}}{A_{\rm loop}}\frac{y}{W} \right) ,
\end{equation} 
where we insist that $\Phi$ is the magnetic flux enclosed in the whole loop rather than the 
flux enclosed in the junction.

To illustrate the magnetic flux modulation of the LDOS, we follow Ref.~\cite{Bretheau2017}
and assume that $A_{\rm loop}/A_{\rm N} = 7.5$ and consider a normal metal of length $L=0.2\xi$ and 
width $W=0.7\xi$. We show in Fig.~\ref{fig-LDOS-loop}(a) the modulation of the energy dependence 
of the LDOS of this junction with the magnetic flux enclosed in the whole superconducting loop. As 
expected, the modulation of the LDOS is progressively suppressed as the magnetic flux increases,
but it does it much more slowly than in the cases shown in Fig.~\ref{fig-LDOS-Bfield} because the 
flux enclosed in the normal region of the junction is much smaller than the total flux enclosed in
the loop (7.5 times smaller). Notice that in the first cycles one can clearly see a hard Andreev gap
(no DOS close to the Fermi energy) that only closes when the total flux is close to a multiple of 
the flux quantum. For completeness, we show in Fig.~\ref{fig-LDOS-loop}(b) the corresponding 
modulation of the supercurrent with the magnetic flux enclosed in the loop. As one can see, 
the current is strongly non-sinusoidal and it decays as the magnetic flux increases. The results
presented here for the LDOS actually resemble those reported in Ref.~\cite{Bretheau2017}
for a S-graphene-S junction, especially for high gate voltages when the graphene Fermi energy is away
from the Dirac point. The main differences are: (i) the experimental results typically show a 
soft gap at low energies, contrary to the hard gap that we obtain, and (ii) the modulation of the 
LDOS in our simulations decays with the magnetic flux more rapidly than in the experiments. In any 
case, it is important to emphasize that we do not aim here at reproducing or explaining the results of 
Ref.~\cite{Bretheau2017} since our model does not incorporate any specific
physics of graphene. Moreover, the authors of that reference estimated a mean free path of 
$l_e \sim 140$ nm and a superconducting coherence length of $\xi\sim 590$ nm, while the junction
length was about 380 nm. This means that those experiments were likely in an intermediate situation
between the clean and the dirty limit. 
\subsection{Relation between the density of states and the Josephson current}

Now, we want to investigate the relation between the DOS and the dc Josephson current.
Let us recall that in a short Josephson junction $(L \ll \xi)$, the whole supercurrent
is carried by the ABSs. In particular, for a single-channel point contact of 
transparency $D$, the ABS energies are given by $E^{\pm}_{\rm A}(\varphi) = 
\pm \Delta \sqrt{1 -D \sin^2 \varphi/2}$. These states carry opposite supercurrents 
\cite{Furusaki1991,Beenakker1991,Bergeret2010}
\begin{equation}
\label{eqn:I_Ea}
I^\pm(\varphi) = \frac{2e}{\hbar} \frac{\partial E^\pm_{\rm A}}{\partial\varphi},
\end{equation}
which are weighted by the occupation of the ABSs. Inspired by this expression,
Bretheau \emph{et al.}\ \cite{Bretheau2017} proposed the following
heuristic formula that relates the Josephson current and the LDOS at zero temperature and 
for a junction of arbitrary length
\begin{equation}
\label{eqn:I}
I(\varphi) =  -\frac{W}{2 \Phi_0} \int_{-\infty}^{\infty} \sign{(-E)} 
J_s(\varphi,v_{\rm F},E,L)dE,
\end{equation}
where $\Phi_0=\hbar/2e$ is the reduced magnetic flux quantum, $W$ is the width of the normal metal, and $J_s$ is the spectral current given by
\begin{equation}
J_s(\varphi,v_{\rm F},E,L) =L \int_{0}^{E} 
\frac{\partial N_n(\varphi,v_{\rm F},\epsilon,L)}{\partial\varphi} d\epsilon,
\label{eqn:Js}
\end{equation} 
where $L$ is the length of the normal metal and $N_n$ is the corresponding LDOS. 
This formula gives exactly Eq.~(\ref{eqn:I_Ea}) whenever we deal with a DOS of the
form,
\begin{equation}
N_n(\varphi,E) = \frac{1}{2} \left[\delta(E-E_A^+(\varphi)) + 
\delta(E-E_A^-(\varphi)) \right],
\end{equation}
which, however, is not always true. Actually, in section \ref{sec-supercurrent} we
proved that Eq.~(\ref{eqn:Js}) is not correct for a single trajectory solution 
by comparing it to the analytical result for a junction of perfect transparency,
see Eq.~(\ref{eqn:js}).
\begin{figure}[t]
\includegraphics[width=0.9\columnwidth,clip]{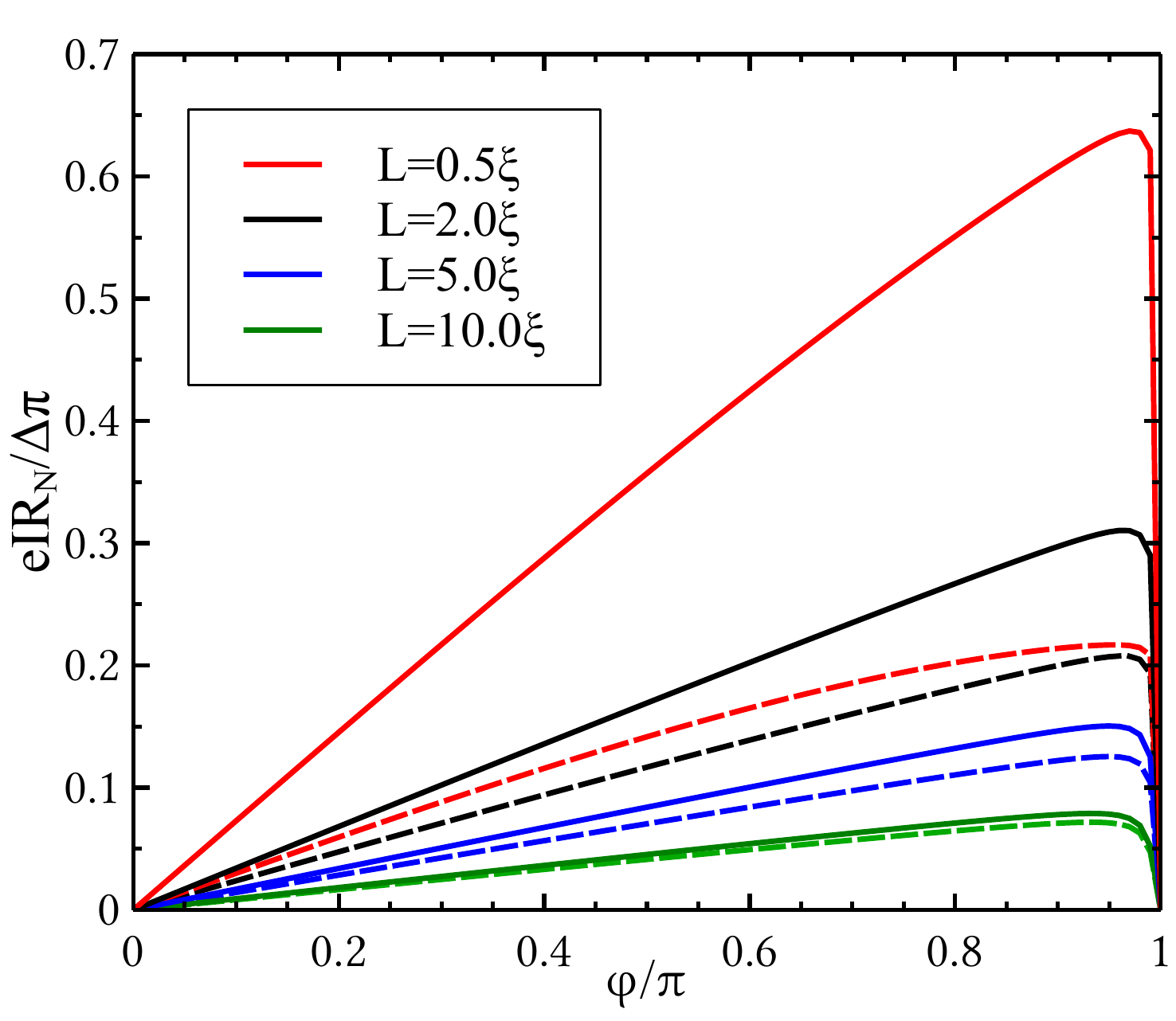}
\caption{(color online) The zero-temperature current-phase relation for a single trajectory 
in a fully transparent junction for various lengths. The solid lines correspond to the exact 
results calculated from Eq.~(\ref{eqn:j_singleTrajecotry}), while the dashed lines
correspond to the heuristic formula of Eq.~(\ref{eqn:I}) proposed in 
Ref.~\cite{Bretheau2017} (dashed lines). The broadening parameter was 
taken $\eta = 0.001\Delta$.}
\label{fig-comparison}
\end{figure}
Here we propose a modified formula based on the global DOS instead. This global 
DOS is defined as an integral of the LDOS over the whole space, which for a 
single-trajectory case (1D) adopts the form
\begin{equation}
N_{\rm total} = \int_{-\infty}^{\infty} N(x) dx = 
\Re{ \bigg[ \int_{-\infty}^{\infty} g(x) dx \bigg] },
\end{equation}
where $N(x)$ is the LDOS along the system. By inserting the single-trajectory 
solution for the Green's function of Eq.~(\ref{eqn:g_solution}) in the previous 
formula and comparing the result with Eq.~(\ref{eqn:js}), one can show that the 
following formula is fulfilled
\begin{equation}
\begin{split}
\label{eqn:js_global}
J_s(\varphi,v_{\rm F},E,L) = & - \int_{0}^{E} d\epsilon \int_{-\infty}^{\infty} 
\frac{\partial N(x,\varphi,v_{\rm F},\epsilon,L)}{\partial\varphi} dx
\\
= &- \int_{0}^{E} \frac{\partial N_{\rm total}(\varphi,v_F,\epsilon,L)}
{\partial\varphi} d\epsilon.
\end{split}
\end{equation}
The minus sign is due to the function $\sign(-E)$ in the Eq.~(\ref{eqn:I}). To illustrate
the difference between this expression and the heuristic formula above, we present in 
Fig.~\ref{fig-comparison} a comparison between our result and the heuristic formula
summarized in Eqs.~(\ref{eqn:I}) and (\ref{eqn:Js}). This comparison is made for a 
single trajectory in a fully transparent junctions and we present results for junctions
of different lengths. As one can see, there are clear deviations between these two 
formulas and they only coincide in the limit of very long junctions. Mathematically,
this can be understood with the help of Eq.~(\ref{eqn:js}). In the limit of 
sufficiently long junctions $L \gg v_{\rm F}/ \sqrt{\Delta^2-\epsilon^2}$ and the
exact formula reduces to the heuristic one. In the opposite limit, i.e., for short
trajectories, the disagreement between both results is quite apparent. Note that 
the normal state resistance that appears in Fig.~\ref{fig-comparison} is defined as
$R_{\rm N} = 2\pi /We^2k_{\rm F}$. 
In order to understand the difference on an analytic level, we can have a look at the local density of states. From Eq.~(\ref{eqn:g_solution}) we can write in the subgap range
\begin{equation}\label{eq:gn}
	g^n(\varphi,E) =  - i \cot\left(\varphi/2-\tilde{\gamma}(E)\right)\,,
\end{equation}
where we defined the energy-dependent phase factor $\tilde{\gamma}(E)=2EL/v_F+\arccos(E/\Delta)$. The Green's function has poles for $\varphi/2-\tilde{\gamma}(E_{Bn})=n\pi$, where $n=0,\pm 1, \pm 2, \ldots$. 
We find the local density of states
\begin{equation}\label{eq:ldos}
	\frac{N(\varphi,E)}{N_0}=\sum_n\frac{\delta(E-E_{Bn}(\varphi))}{\Delta L/v_F+1/\sqrt{1-E_{Bn}^2(\varphi)/\Delta^2}}\;.
\end{equation}
The difference to the global density of states (only the $\delta$-functions) is related to the leakage of Andreev states into the superconductor, which strongly depends on the energy of the bound state (and hence on the phase).
The phase dependence is most striking in the short junction limit. Here we obtain a single bound state at $E_B(\varphi)=\Delta\cos(\varphi/2)$ and therefore
\begin{equation}\label{eq:ldosshort}
	\frac{N(\varphi,E)}{N_0}=\sin(\varphi/2)\delta(E-E_{B}(\varphi))\,.
\end{equation}
Obviously, the local density of states differs drastically from the simple $\delta$-function and this explains the deviations from the heuristic formula and the full result illustrated in Fig.~\ref{fig-comparison}. 
As final remark on the heuristic formula, we note that we have checked numerically that our relation (\ref{eqn:js_global}) for junctions of finite transparency works correctly as well. In particular in the limit of zero length, one obtains in  fact  $N(\varphi,E)/N_0=\sqrt{D}\sin(\varphi/2)\delta(E-E_{B}(\varphi))$, where $E_B(\varphi)=\Delta\sqrt{1-D\sin^2(\varphi/2)}$.
Hence, one has to be careful, when extracting the spectral supercurrent density and the current-phase relation from local tunneling measurements.

\section{Conclusions} \label{conclusion}

With the goal to help to interpret future experiments, we have presented here 
a comprehensive theoretical study of the LDOS in clean 2D SNS junctions. Making
use of the quasiclassical Green's function formalism, we have calculated both the
LDOS and the Josephson current as a function of parameters such as the
length and the width of the junction, the transparency of the system, and we have
studied the role of a weak magnetic field. 

First, we have shown how discrete ABSs become visible inside the gap for short junctions. At finite reflectivity $R$ a phase-independent minigap $\sim\sqrt{R}\Delta$ is present, but the LDOS still reflects the energies of the Andreev bound states. The phase dependence above the gap is rather weak and further decreased by a finite reflection.

Next, we have studied the effect of a finite length of the junction. A finite reflection still leads to a minigap, but this diminishes for longer junction. Finally the spectrum of a long junction with the linear phase-dependent Andreev states is emerging. 

The effect of a magnetic field leads to a rather complex behavior. Interfering trajectories due to the finite flux lead to a vanishing phase-dependence of the density of states. This is in analogy to usual Fraunhofer suppression of the Josephson critical current for a magnetic flux threading the junction. Generically, we observe that for ballistic transport, the minigap closes at a phase-difference of $\pi$ and reopens for a finite flux. At large fluxes there is no gap anymore.

To make a connection to the experiment of Bretheau and coworkers \cite{Bretheau2017}, we have studied the experimental setup, for which the phase difference is imposed by an additional loop and a magnetic field leads to phase bias simultaneously to a flux threading the junction. As a result we qualitatively reproduce the LDOS pattern, but find, surprisingly a stronger suppression with magnetic field compared to experiment. We attribute this to a possible inhomogeneous current distribution in the experiment caused by local perturbations.

Finally, we have investigated the relation between the LDOS and the Josephson current proposed in 
Ref.~\cite{Bretheau2017}. We have shown that, in general, it has to be modified because the localization of the bound states in the normal region strongly depends on phase. We propose a new relation, which takes this effect into account and will have important implication for future experiments. Unfortunately, the relation is less universal and requires a more sophisticated modelling by theory. This is most likely even worse in the presence of impurities.

\section{Acknowledgement} \label{Acknowledgement}

We thanks L. Bretheau for numerous discussions on the experimental results of 
Ref.~\cite{Bretheau2017}. D.N. and W.B. have received funding from the 
European Union’s Horizon 2020 Research and Innovation Programme under the Marie Skłodowska-Curie 
grant agreement No 766025 and the DFG through SFB 767. J.C.C. thanks MINECO (contract No.\ FIS2017-84057-P) for financial support as well as the DFG through SFB 767 for sponsoring his stay at the University of Konstanz 
as Mercator Fellow. 


\end{document}